\documentclass[12pt]{iopart}

\usepackage{setstack}
\usepackage{graphicx}

\begin{document}

\title[Real Time Diagnostics and Feedback Algorithms for JET]
       {Development of Real Time Diagnostics and Feedback 
       Algorithms for JET in view of the Next Step}

\author{A.Murari\dag\ \footnote[3]{corresponding author \underline{andrea.murari@igi.cnr.it}}, E.Joffrin\ddag, R.Felton\S, 
        D.Mazon\ddag, L.Zabeo\S, R.Albanese\P, P.Arena\pounds, G. Ambrosino$\hbar$, M. Ariola$\hbar$, O.Barana\dag, 
        M.Bruno\pounds, L.Laborde\ddag, D.Moreau\ddag, F.Piccolo\S, F.Sartori\S, F.Crisanti$\sharp$, 
        E. de la Luna$\flat$, J.Sanchez$\flat$ and EFDA-JET Contributors}

\address{\dag\  Consorzio RFX – Associazione EURATOM ENEA per la Fusione, Corso Stati Uniti 4, I-35127, Padua, Italy}
\address{\ddag\  Association EURATOM-CEA, CEA Cadarache, 13108 Saint-Paul-lez-Durance, France}
\address{\S  Euratom/UKAEA Fusion Assoc., Culham Science Centre, Abingdon, Oxon, OX14 3DB, UK}
\address{\P  Assoc. Euratom-ENEA-CREATE, Univ. Mediterranea RC, Loc. Feo di Vito, I-89060, RC, Italy}
\address{\pounds  Assoc. Euratom-ENEA-CREATE, Univ.di Catania, Italy}
\address{$\hbar$  Assoc. Euratom-ENEA-CREATE, Univ. Napoli Federico II, Via Claudio 21, I-80125 Napoli, Italy}
\address{$\sharp$ Associazone EURATOM ENEA sulla Fusione, C.R. Frascati, Italy}
\address{$\flat$  Associacion EURATOM CIEMAT para Fusion, Avenida Complutense 22, E-28040 Madrid, Spain}

\begin{abstract}

Real time control of many plasma parameters will be an essential aspect in the development 
of reliable high performance operation of Next Step Tokamaks. The main prerequisites for 
any feedback scheme are the precise real-time determination of the quantities to be controlled, 
requiring top quality and highly reliable diagnostics, and the availability of robust control algorithms. 

A new set of real time diagnostics was recently implemented on JET to prove the feasibility 
of determining, with high accuracy and time resolution, the most important plasma quantities. 
With regard to feedback algorithms, new model-based controllers were developed to allow a more 
robust control of several plasma parameters.  Both diagnostics and algorithms were successfully 
used in several experiments, ranging from H-mode plasmas to configuration with ITBs.

Since elaboration of computationally heavy measurements is often required, significant attention 
was devoted to non-algorithmic methods like Digital or Cellular Neural/Nonlinear Networks. 

The real time hardware and software adopted architectures are also described with particular 
attention to their relevance to ITER.

\end{abstract}

\maketitle

\section{\label{sec:one} Introduction}

In the last years, the fusion community has witnessed a significant proliferation in the 
number of real time control experiments. This is due to a variety of reasons, among 
which the most important are the increased sophistication of the scenarios and the 
advanced physical issues attacked \cite{Watkins}. With regard to the scenarios, the Elmy H mode, 
in the continuous attempts to increase the performances moving closer to the 
Greenwald limit, requires high elongations and triangularity, with the related 
difficulties in terms of stability of the plasma column. The so-called advanced 
scenarios, in their turn, rely every day more on involved profiles of both pressure and 
current, posing not negligible challenges to the control systems. The need to 
understand many unresolved issues, like the physics of ELMs or the formation and 
sustainment of ITBs, also contributes to the variety of requests for sophisticated 
feedback schemes. In JET the demand for more advanced real time control is 
particularly felt due to the extremely wide scope of its research program and to the 
more acute safety implications of a big device \cite{Joffrin}. As a consequence a broad range of 
different feedback schemes has been recently developed covering a wide range of 
tasks from simple event driven actions to multivariable, distributed control of profiles.

In Nuclear Fusion, as in any other field, the first step for a successful feedback is the 
proper identification of the system to be controlled. This means that the relevant 
parameters of the system must be measured with adequate accuracy, reliability and 
speed. In the case of Tokamak plasmas, the responsibility of the identification falls 
mainly on plasma diagnostics. In the last years, the main drive behind the success of 
JET feedback control experiments were the major improvements in the number and 
reliability of the measurements \cite{Felton}. The signals available in real time at JET now 
cover all relevant parameters, ranging from the magnetic configuration to the kinetic 
quantities. In addition to the traditional plasma shape obtained form the pick-up coils 
and the flux loops, the $q$ profile is derived from the Faraday rotation measurements 
\cite{Zabeo}. Both the electron and the ion fluid are diagnosed. The electron density profile is 
provided by the LIDAR \cite{Felton} and the interferometer, whereas the ECE \cite{Riva} guarantees 
an electron temperature profile with much higher time resolution than the Thomson 
Scattering. Active Charge exchange Recombination Spectroscopy (CXRS) is of 
course the main system to derive real time information, temperature and velocity 
profiles, of the ion fluid \cite{Heesterman}. 

A series of validated codes are also routinely used to obtain derived quantities. The 
internal plasma inductance $l_i$ and the derived confinement parameters are calculated 
from the Shafranov integrals \cite{Barana}. A new shape controller, called extreme Shape 
Controller (XSC) \cite{Albanese} has already been used in highly shaped configurations, whereas 
an optimised equilibrium code EQUINOX \cite{Bosak} solves the Grad-Shafranov equation, 
taking into account also internal magnetic measurements, like polarimetry, on a finite 
element mesh instead of a regular grid like EFIT. 

These new tools were extensively used in the last years and they contributed 
significantly to the scientific program of the entire experiment. For example they 
constituted an indispensable prerequisite for some of the most ambitious feedback 
programs, like the simultaneous control of current and density profiles. These control 
schemes made extensive use also of JET actuators, which toroidal, poloidal and 
divertor coils, gas fuelling, neutral beam injection and Radio Frequency (ICRH) and 
Lower Hybrid (LH) waves.
 
JET real-time system is also a good environment to test innovative computational 
concepts, both software, like Digital Neural Networks (DNN) \cite{Barana2}, and hardware, like 
the new chip technology of Cellular Neural/Nonlinear Networks (CNN) \cite{Arena}. The 
information and experience matured at JET in years of real time developments 
provides also a good basis in the perspective of ITER. JET architecture and general 
approach, implementing a distributed system, could be successfully translated to 
ITER. On the other hand, on the route to the next step, the validation of new 
measurement techniques for some physical quantities of reactor relevance is still 
required.

With regard to the structure of the paper, the architecture of JET real time control and 
the main diagnostics and codes available in real time are reported in section \ref{sec:two}. 
Some of the most recent and interesting feedback experiments, heavily relying on the new 
real time diagnostics and algorithms, are described in section \ref{sec:three} and \ref{sec:four}. 
More advanced approaches, still under development, for real time elaboration of diagnostic 
data, involving soft computing and hardware neural Networks, are the subject of 
section \ref{sec:five}. The main problems to be faced in developing diagnostic concepts for ITER 
a reviewed in section \ref{sec:six}.

\section{\label{sec:two} Architecture of JET real time control system and diagnostics}

JET real time control implements a distributed system, with many independent 
stations, communicating via an ATM protocol \cite{Felton2}. This multi-platform approach 
(PCI, VME etc) offers several advantages, with respect to the mainframe, centralised 
solutions, which were more popular in the past also in the field of Nuclear Fusion. 
One of the main characteristics of the present architecture is its great flexibility, which 
is essential in such a fast evolving field, with diagnostics and feedback algorithms 
being continuously added or upgraded. The potential of the adopted solution to 
implement parallel computing is also a very important feature, which is not to be 
neglected given the fast time response of many plasma phenomena to be controlled 
and the quantity of data to be processed in next step machines. It is therefore strongly 
believed that JET approach should be considered a very good reference for ITER real 
time control system.

The flexible and adaptive architecture of JET control system has allowed including 
many new diagnostics in the real time project in a very efficient way. Now the vast 
majority of the most relevant measurements, from the equilibrium to the confinement 
parameters, are routinely available in real time \cite{Felton}. Also kinetic and profile quantities 
are provided with more than satisfactory time and space resolution, as can be seen 
from the summary table \ref{tab:diagnostics}.

\begin{table}
 \caption{Main real time signals and derived quantities routinely 
                                 available in real time at JET}
 \begin{indented}
  \label{tab:diagnostics}
 \item[]\begin{tabular}{@{}llll}
 \br
  Physics & Diagnostic & Size & Cycle (ms)\\
 \mr
  $T_e$(R)       & ECE         & 48(96) & 5\\ 
  $ITB_e$(R)     & EXE         & 48     & 5\\ 
  $T_e$(R)       & LIDAR       & 50     & 250\\
  $N_e$(R)       & LIDAR       & 50     & 250\\
  $T_i$(R)       & CX          & 14     & 50\\
  $V_{rot}$(R)     & CX          & 14     & 50\\
  $ITB_i$(R)     & CX          & 14     & 50\\
  $\gamma$(R)  & MSE         & 25     & 2\\
  LID          & FIR         & 8      & 2\\
	FAR	         & FIR         & 8      & 2\\
  LCFS				 & XLOC  		   & 100    & 2\\
  $\beta$,$l_i$ & Confinement & 20     & 2\\
  Flux         & EQX         & 100    & 25\\
  q(r/a)	     & FIR / XLOC  & 10     & 2\\
  q(r/a)       & MSE / EQX   & 10     & 25\\
	$ITB_e$(r/a)   & ECE / EQX   & 10     & 25\\
  $ITB_i$(r/a)   & CX / EQX    & 10     & 25\\
  Rad'n 		   & Bolometer	 & 48	 	  & 5\\
	Imp'y        & VUV         & 8      & 20\\
  Imp'y				 & Vis.        & 16		  & 20\\
	ELM          & Vis.        & 3*3    & 100\\
	H:D:T				 & Vis.        & 4*3    & 20\\
  $T_i$ core     & X-ray       & 8			& 20\\
							 &             &        &   \\
  Ipla				 & Magnetics   & 1			& Analog\\
	MHD n=1      & Magnetics   & 1      & Analog\\
	MHD n=2			 & Magnetics   & 1      & Analog\\
	RNT				   & Neutronics  & 1      & Analog\\
	Hard Xray	   & Neutronics	 & 1      & Analog\\
	Density			 & FIR			   & 1			& Analog\\
 \br
 \end{tabular}
 \end{indented}
\end{table}

From table \ref{tab:diagnostics} it is very apparent the remarkable progress of JET diagnostics in the 
direction of the real time, which is now not limited to the traditional magnetic 
measurements for plasma positioning and control. The electron fluid is nowadays 
quite well diagnosed, since the temperature is given by the ECE and the density can 
be obtained from both the interferometer and the LIDAR Thomson Scattering.  The 
ECE radiometer comprises 96 tuned heterodyne microwave receivers covering the 
much of the radial extent of the plasma, for most toroidal fields. The real time 
algorithm acquires the 96 signals and, after proper filtering, applies calibration 
constants derived from comparison with the absolute-reading ECE Michelson Fourier 
Interferometer, providing profiles with $5ms$ time resolution. For the LIDAR system, 
the real time approach consists of fitting to pre-calculated intensities, depending on 
electron density and temperature, the backscatter echo from the plasma. The system 
processes the data in the almost same way as the inter-shot code in less than $10ms$, 
resulting in 50 point profiles at $4Hz$ (due to the repetition rate of the laser $250ms$). 
The electron cycloctron emission data allows in its turn the determination of the $\rho^*T$ 
profile \cite{Laborde}. On the basis of the electron density, obtained by inversion of the 
interfometric measurements, the safety factor profile can also be calculated in real 
time, using the flux surface topology obtained by the pick up coils \cite{Zabeo}. The last years 
have also witnessed a remarkable progress in diagnosing the ion fluid, whose 
temperature and velocity are now routinely provided using Charge Exchange 
Recombination Spectroscopy \cite{Heesterman}. In particular the toroidal velocity is considered 
particularly relevant and it could be exploited much further in the future for very 
interesting feedback schemes, like the control of ion ITBs.

The shown performances have been obtained thanks to significant improvements both 
in the hardware and the software. In the last years many diagnostics have become 
more reliable and communication technologies has also witnessed dramatic steps 
forward. From the point of view of data analysis, providing a quantity in real time 
mainly implies very often a critical revision of the off-line algorithms, to find a trade 
off between accuracy and time resolution. Proper approximations, linearization of 
quantities and adoption of robust fitting routines, together with careful software 
engineering, are the main ingredient, which normally can grant the desired results, 
both in terms of accuracy and time resolution.

It is also worth noting that not only a quite comprehensive set of signals is available in 
real time at JET but also some fundamental derived quantities are calculated, by 
optimised and reliable codes. Particularly interesting are all the basic confinement 
quantities, obtained from the Shafranov integrals \cite{Barana}, and the magnetic equilibrium 
from the magnetic measurements (EQUINOX code) \cite{Bosak}, as again reported in table \ref{tab:diagnostics}. 
In the case of these codes, a reasonable compromise between the accuracy of the 
derived quantities and the computational efforts is again an essential ingredient of a 
successful strategy for providing satisfactory and reliable information. This was 
already achieved in the case of EQUINOX making use only of the external coils and 
now the attention is concentrating on finalising the version of the code capable of 
accommodating also the internal measurements (polarimetry and MSE).

\section{\label{sec:three} Control of highly shaped plasmas}

As mentioned in the introduction, the main drives behind the development of real time 
diagnostics at JET are the requirements of the scenarios and the advanced physics. 
Recently the research on the Elmy H mode has moved in the direction of producing 
plasmas with increased elongation and triangularity. These strongly shaped 
configurations are quite vulnerable to significant deformations of the shape in 
presence of strong variations of $\beta_{pol}$ and/or the internal inductance $l_i$. In this 
framework, a new controller, called the eXtreme Shape Controller (XSC) \cite{Albanese} was 
explicitly designed at tested to improve the control of these highly shaped plasmas. In 
JET 8 actuators, namely 8 Poloidal Coils, are available to control the plasma shape, 
which is described in terms of a set of geometrical descriptors (GAPs). These GAPs 
are the distance of the last closed flux surface from the first wall along predefined 
directions. They are obtained from the magnetic measurements of fields, fluxes and 
flux differences by standard analysis methods. The previous shape controller (SC) 
was conceived to perform the feedback control on each of the 8 actuators by using as 
inputs to the system either the currents flowing into the Poloidal Circuits (current 
control) or a limited number of the actual measured GAPs (GAP Control). The new 
XSC receives the errors on 38 indicators of the plasma shape (32 GAPs plus the two 
coordinates of the X and the two strike points) and calculates the "smallest" currents 
needed to minimize the error on the "overall" shape in a least square sense.

The design of the XSC for JET single-null configurations is based on a linearised 
plasma model approach, implemented by the CREATE-L and CREATE-NL codes 
\cite{Ambrosino}. These plasma-modeling tools were specifically adjusted for JET topology, 
taking into account both the iron core and the eddy currents induced in the passive 
structures. 

\begin{figure}[ht]
 \centering
 \includegraphics[scale = 0.6]{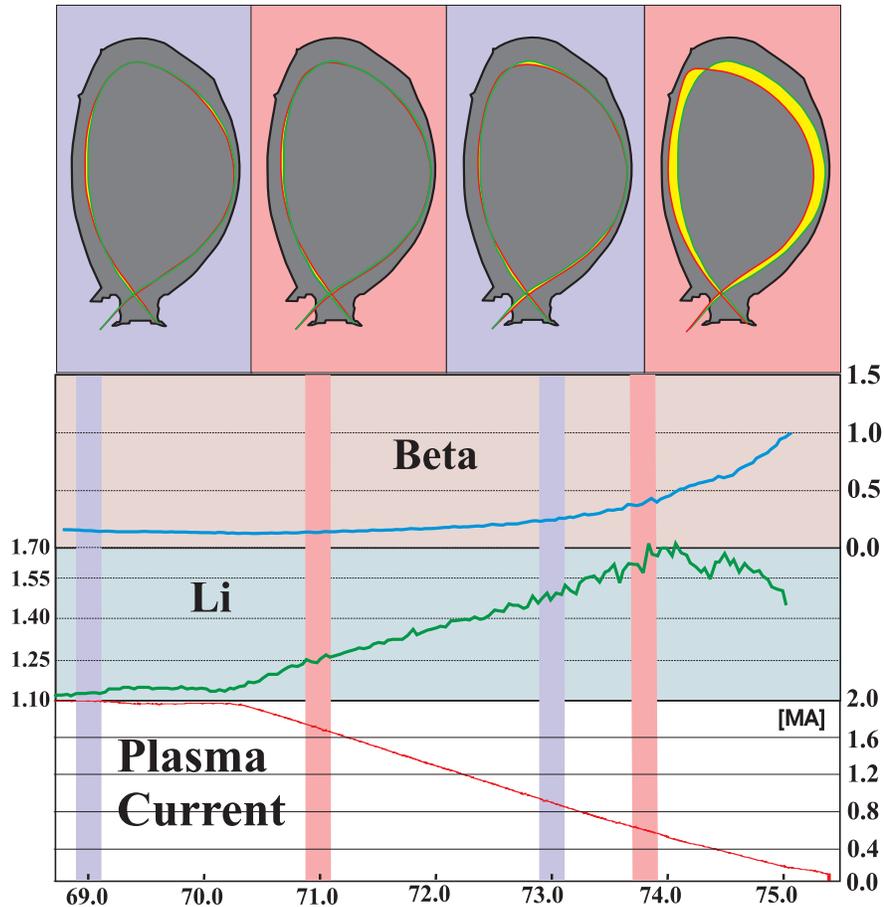}
 \caption{Performance of the XSC in the case of very significant variations of $\beta_p$, $l_i$ and Ip. 
   The yellow region indicates the distance between the target shape and the one really achieved 
   by the controller}
    \label{fig:one}
\end{figure}

With regard to the controller implementation, the chosen approach identifies the 
principal directions of the algebraic mapping between coil currents and geometrical 
descriptors using the singular value decomposition (SVD). These principal directions 
can be translated into 8 linear combinations of currents, which represent one linear 
combination of geometrical descriptors each. Such an approach allows solving the 
original multivariable control problem using a set of separate PID controllers. To 
alleviate the burden on the actuators, the SVD orders the principal directions as a 
function of the current to shape sensitivity and normally only the first 5 or 6 directions 
(out of 8) are used. The control algorithm is optimized to obtain the most efficient 
distribution of the control currents, compromising between the effort of the actuators 
and the tracking error on the plasma shape. 

As a consequence of this different approach, the XSC manages to achieve the desired 
shape with typically an average error of about $1 cm$ on the 48 descriptors. An example 
of the capability of the XSC is reported in figure \ref{fig:one}, which shows the difference 
between the desired and obtained shape for a quite extreme situation a the end of a 
JET discharge. The controller manages to keep the shape more or less constant even 
in the presence of large variations of $\beta_p$, $l_i$ and also Ip. In general the XSC has already 
been tested successfully for variations $\Delta l_i$ up to $0.5$ and $\Delta\beta_p$ up to $1.5$.

\section{\label{sec:four} Control of profiles and ITBs in advanced scenarios}

A linearisation approach has been adopted also for the control of the current and 
pressure profiles. The long term objective of this program consists of being able to 
sustain internal transport barriers (ITB) in high performance plasmas, with a large 
bootstrap current fraction and possibly to reach steady state operation ('advanced 
tokamak' program). In the case of these "advanced tokamak scenarios", the 
challenges to the control become particularly severe because the non-linear coupling 
between the pressure and current profiles is particularly involved, given the relevant 
fraction of bootstrap currents and the presence of ITBs. In order to have reasonable 
chances of success the plasma must be controlled on the time scale of both the current 
diffusion and the thermal evolution. Moreover, the adopted approach must preserve 
the distributed nature of the problem, because accurate control of the profiles must be 
achieved in order to properly influence the barriers. A linearised, model based, 
distributed control system was therefore adopted for the simultaneous control of the $q$ 
and $\rho^*T$ profiles \cite{Moreau}. 

The objective of the experiments reported in this paper consisted of 
demonstrating for the first time the feasibility of simultaneous combined control of 
the current and electron pressure profiles in presence of internal transport barriers. 
This was obtained with the distributed-parameter version of the
theoretical method, in which the spatial current and pressure profiles where
described by a suitable set of basis functions, 5 cubic splines and 3
piecewise-linear functions, respectively \cite{Laborde}.The designed multiple-input-multiple-output 
(MIMO) controller operates all the three available heating and current 
drive actuators (NBI, LHCD and ICRF) during the high power phase of the discharge. 
The chosen scenario was a typical reversed shear configuration obtained with $2.5MW$ 
LHCD in the preheat phase during which the plasma current was ramped up to $1.7MA$, 
at a line integrated plasma density about $3\times10^{19} m^{-2}$. The determination of the 
steady-state responses to variations of the heating and current drive powers was 
obtained from the analysis of four dedicated open loop discharges \cite{Laborde}. 

\begin{figure}[ht]
 \centering
 \includegraphics[scale = 0.5]{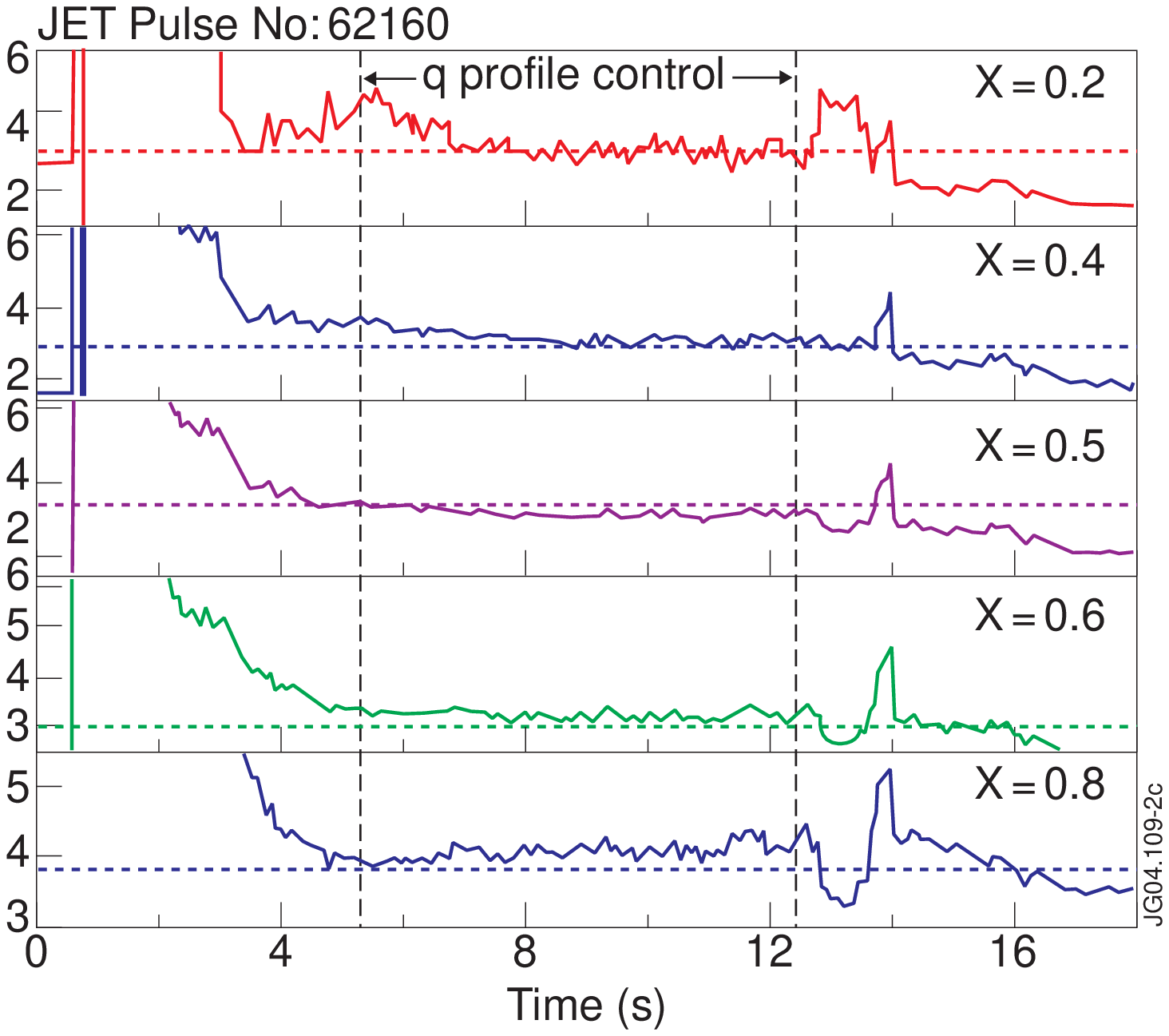}
 \includegraphics[scale = 0.48]{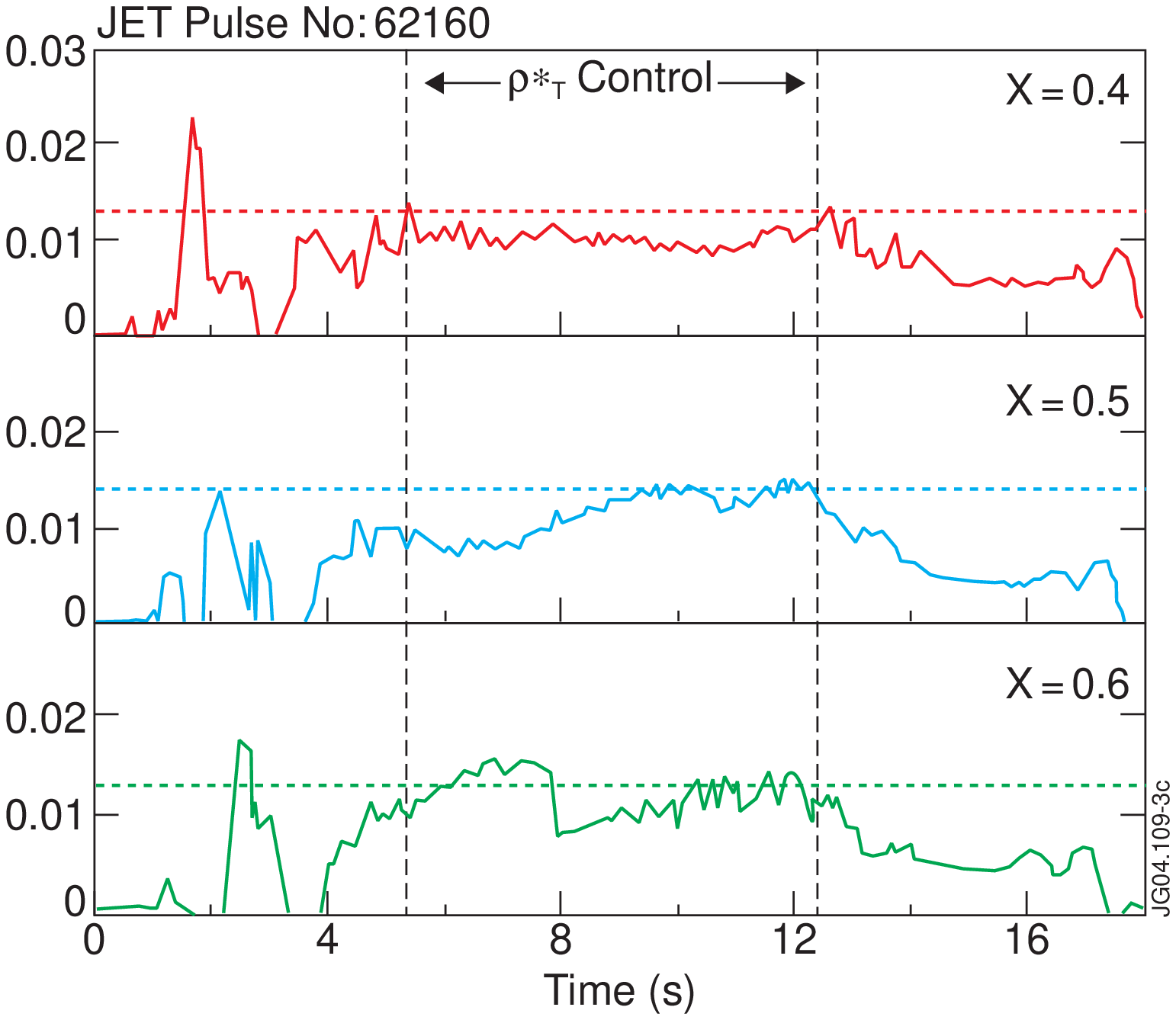}
 \caption{Time evolution of $q$ toward the set points during the control phase. 
Right: Time evolution of $\rho^*T$ toward the set points during the control phase}
 \label{fig:two}
\end{figure}

\begin{figure}[ht]
 \centering
 \includegraphics[scale = 0.8]{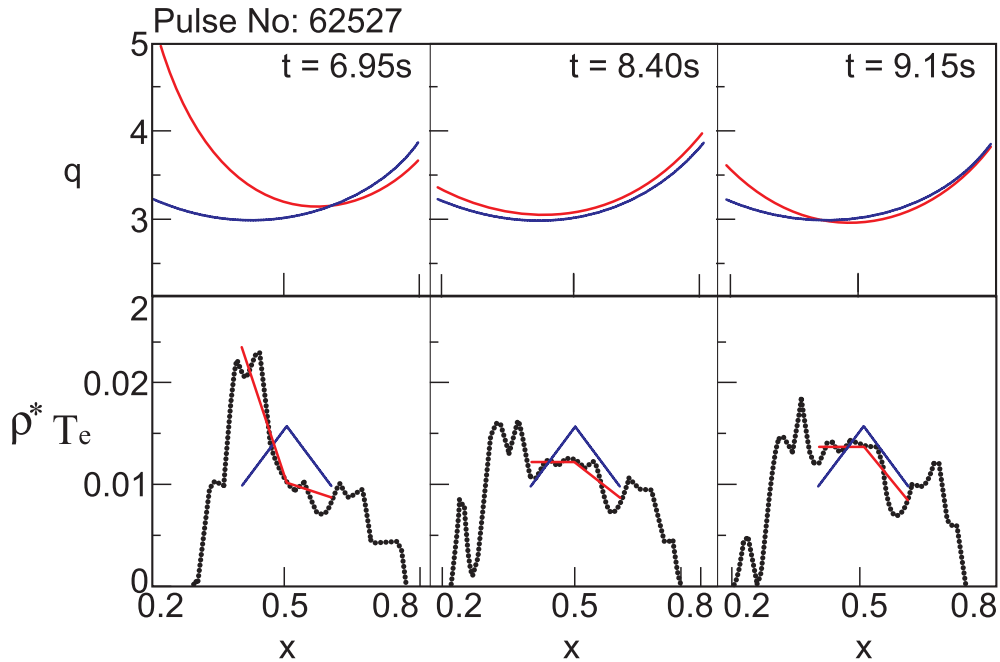}
 \caption{Control of a weak barrier in a reversed $q$ profile discharge. 
The controller succeeds in both counteracting the current diffusion and moving the barrier 
in the more external region of the plasma}
\label{fig:three}
\end{figure}

In figure \ref{fig:two}a the time evolution of $q$ at the five controlled points is reported, to show 
how the target values are properly achieved. The evolution of the $\rho^*T$ profile in the 
controlled region is reported in figure \ref{fig:two}b, from which it can be seen how the 
controller manages to force the plasma also toward this request, in parallel with the 
control of the current. A more intuitive representation of the relevant physical 
quantities is reported in figure \ref{fig:three} for the shot number 62527, in which a weak barrier 
was controlled in real time with a reversed shear profile. It must be mentioned that the 
control was achieved on time scales much shorter than the local resistive time, mainly 
due to the limitations of the actuators. So, even the profile control was demonstrated 
for the first time, the robustness of the approach will have to be confirmed by longer 
duration pulses.

The obtained stabilisation of a barrier for almost the entire shot and with two different 
$q$ profiles is a very interesting result, which could have various applications in JET 
future experimental program.  The robustness of the controller, with regard to ELMs 
and strong MHD activity is also an important aspect for JET and in ITER perspective.

\section{\label{sec:five} Advanced computational techniques and technologies for real time control}

The feedback schemes illustrated in the previous examples, even if they make use of 
sophisticated diagnostics and advance control algorithm, are all based on the standard 
approach of linearisation. Moreover, to guarantee the necessary reliability, they also 
tend to rely on commercial technology. On the other hand, in many fields a lot of 
progress has been recently made in real time computational concepts and components 
and therefore more innovative solutions can be envisaged. In the present section, 
results of some new approaches using Digital Neural Networks (DNN) \cite{Barana2} and 
Cellular Neural/Nonlinear Networks (CNN) \cite{Arena} are presented, to illustrate the 
potential applications of recent software developments and hardware technologies. 
These innovative computational approaches were applied to particularly difficult 
problems like tomographic reconstructions and fast image processing.

From the point of view of data analysis, tomographic reconstructions are considered a 
quite difficult issue in tokamak plasmas. In general tomographic inversions are ill 
posed problems, in the sense the more than one solution is compatible with the 
experimental data and this difficulty is strongly aggravated by the poor accessibility 
of fusion machines.  Moreover, given the topology of the emission in JET, the relation 
between the total radiated power and the line-integrated measurements is a non-linear 
one. Therefore, due also to the computational complexity of the task, in order to 
obtain the total radiated power and the power emitted in the divertor in real time, it 
was decided to try specifically designed DNNs and train them using the total emitted 
power derived from the tomographic reconstrunctions. A multilayer perceptron, with 
one layer of hidden units trained with an error back-propagation learning algorithm, 
resulted more than adequate to the task. For the activation function a sigmoid was 
chosen, to make the DNNs nonlinear transfer functions. In addition to 28 bolometric 
chords also three geometrical factors (elongation, upper and lower triangularities) 
were included in the set of inputs. The training set included about 2700 patterns for 
the divertor configuration and 250 patterns for the limiter configuration. The 
percentage of the DNN estimates that fall in the  $\pm$20\% intervals, centered on the total 
radiation calculated with the tomographic reconstruction, is more than 90\%.  Within 
an interval of $\pm$10\% with respect to the tomographic inversion, which is a value 
comparable with the error bars of the method, fall almost 85\% of the DNN estimates 
and comparable results are obtained for the evaluation of the radiation emitted in the 
divertor region.

\begin{figure}[ht]
 \centering
	\includegraphics[scale = 1]{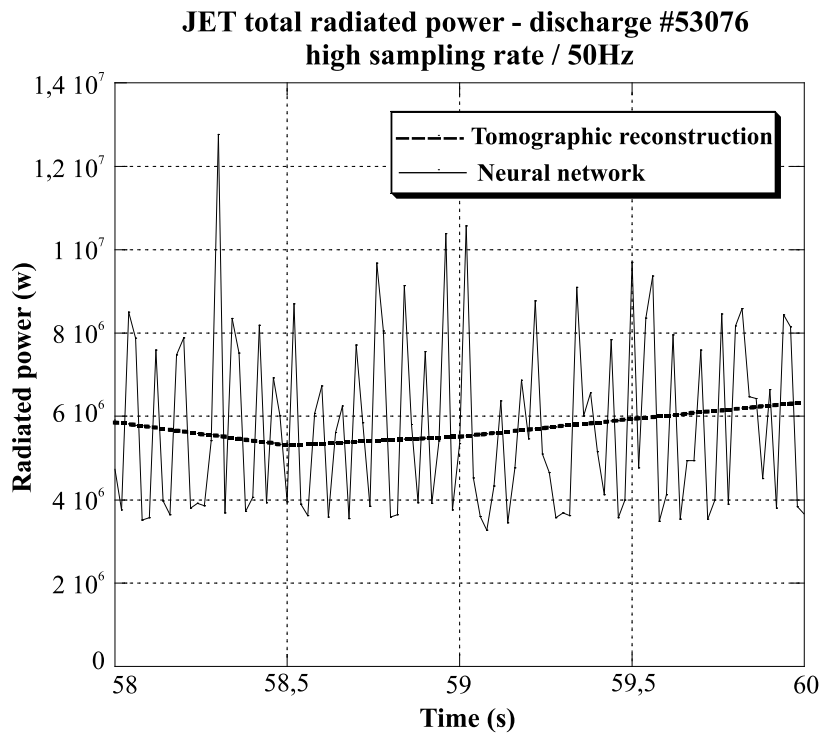}
 \caption{NN estimate of the total radiated power during ELMs}
  \label{fig:four} 
\end{figure}

In addition to the accuracy, also the generalization capability of the DNNs should be 
emphasized. As shown in figure \ref{fig:four}, the designed DNN is capable of following the 
evolution of the total radiated power during ELMs, even if ELMs were not included 
in its training set. This constitutes a quantitative prove of the more general, even if 
qualitative result, i.e. that DNNs perform better than other possible linear methods 
particularly in the case of unusual and unforeseen situations, a fact which of course 
could be of great relevance in the perspective of ITER. 

In current-generation tokamaks, the plasma shape is of primary importance not only 
for protection and control but also for achieving better fusion performance. The 
accurate localisation of the strike points on the divertor plates is essential to estimate 
the power load on the protection tiles, which affects the recycling properties of the 
configuration and bears strong consequences on operation safety. For many years, at 
JET the magnetic reconstruction of the separatrix done by the XLOC code has 
provided quite accurate and robust results. On the other hand, a known weakness of 
the magnetic information is its vulnerability to non-contemplated deviations caused 
by eddy currents induced in the metallic structures by fast transients. Moreover, in the 
perspective of ITER, the long pulse mode of operation raises several questions about 
the stability of the magnetic measurements, which have not been tackled yet. 

A possible alternative and/or support to the magnetic reconstruction approach could 
reside in the use of visual information, for example to identify the position of the 
strike points. In this perspective significant amount of work has been recently devoted 
to developing a technology capable of providing the localisation of the strike points 
from 2D sensors with adequate time resolution. Indeed one of the main difficulties of 
image processing for these applications is typically the need to obtain the required 
output on a millisecond time scale to follow fast phenomena like the ELMs. To meet 
these requirements, great attention has been devoted to the CNN technology. CNNs 
are two dimensional arrays of simple, identical, locally interconnected nonlinear 
dynamic circuits, called cells. These cells are arranged in a rectangular grid where 
each cell interacts with its nearest neighbours. In this way the CNN can implement 
suitable fast algorithms for image processing. The version of the chip tested at JET is 
a new generation 128x128 Focal-Plane Analog Programmable Array Processor 
(FPAPAP). Manufactured in a $0.35\mu m$ standard digital 1P-5M CMOS Technology. 
The chip, identifired by the acronym ACE16K, contains about four millions 
transistors, 80\% of them working in analog mode, with relatively low power 
consumption ($<4W$, i.e. less than $1\mu W$ per transistor). The heart of the chip is an array 
of 128x128 identical, locally interacting, analog processing units designed for high 
speed image processing tasks requiring moderate accuracy (around 8bits). 

\begin{figure}[ht]
 \centering
	\includegraphics[scale = 1.3]{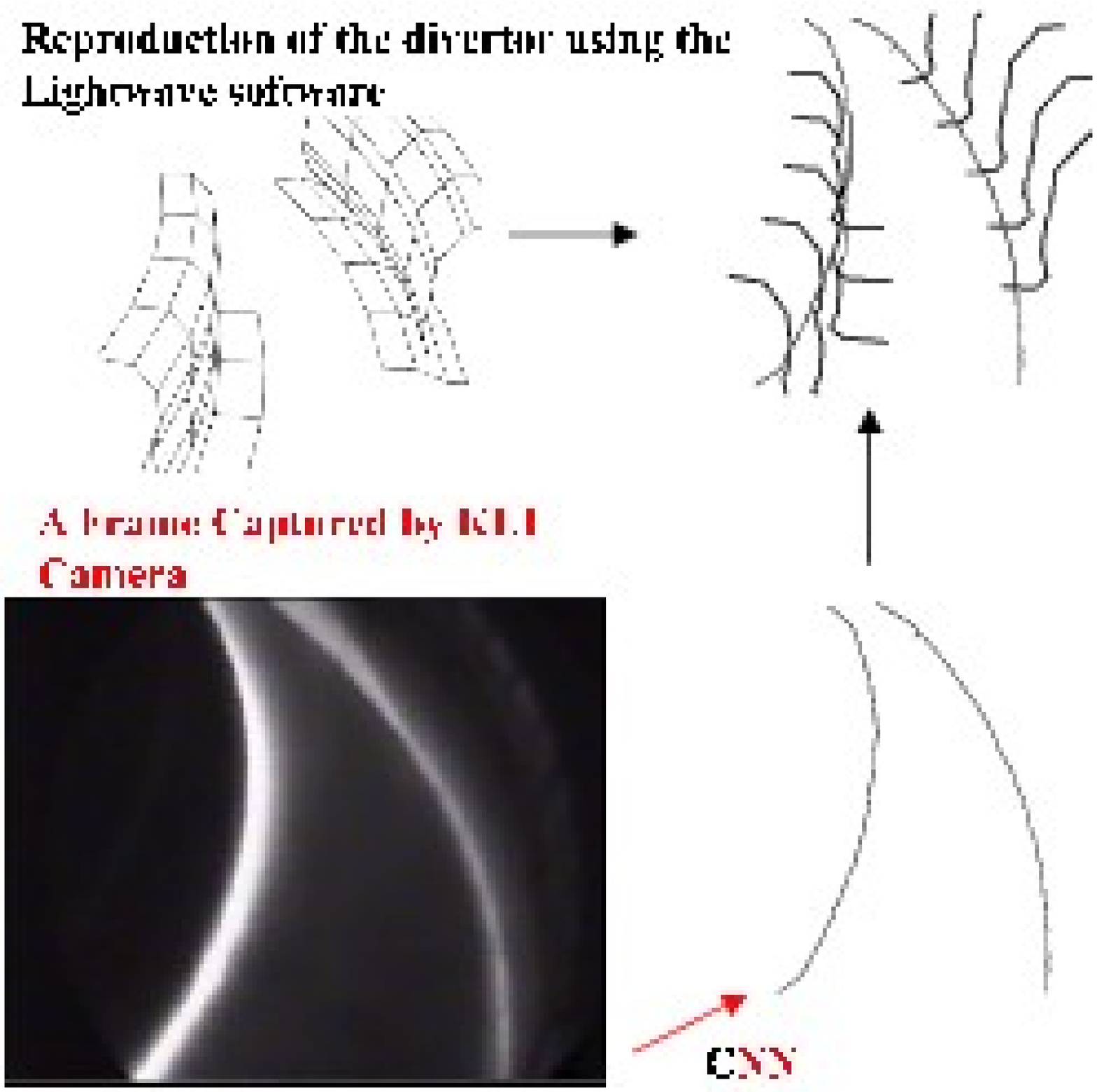} 
\caption{The approach used to determine the strike point position with the CNN}
\label{fig:five} 
\end{figure}

Although ACE16K is essentially an analog processor (computation is carried out in 
the analog domain), it can be operated in a fully digital environment. For this purpose, 
the prototype incorporates a bank of Digital-to-Analog (for input) and Analog-to-
Digital (for output) converters at the images I/O port. ACE16K is conceived for two 
alternative modes of operation. First, in applications where the images to be processed 
are directly acquired by the optical input module of the chip, and second, as a 
conventional image co-processor working in parallel with a digital hosting system that 
provides and receives the images in electrical form. This second operational mode is 
the only one tested so far at JET.  The images of JET CCD visible camera (KL1 
diagnostic, dynamical range of 8 bits, CCD-chip with 751x582 pixels), viewing the 
divertor in a nearly tangential geometry, were used as input (see fig.\ref{fig:five}). The 
information representative of the Strike point position can be derived by the 
brightness of the image. A suitable algorithm has been devised to extract the co-
ordinates of the strike points from the camera images. A typical result is reported in 
figure \ref{fig:five}.

The CNNs guarantee high speed and extreme accuracy in the detection of strong 
features in images, like the position of the strike points. The prospects of the image 
processing to capture the strike points look very promising. At the moment the main 
research is in the direction of investigating the maximum speed of the chip. With 
regard to the algorithms for the identification of the strike points, the preliminary 
comparison of the CNN results with the estimated of XLOC based on the magnetic 
measurements is more than satisfactory. The application of the same approach to 
infrared cameras, instead of the usual CCD for the visible, is also considered 
relatively straightforward, since the pixels of these sensors are normally read with 
CMOS circuits. In the reactor perspective, the main weakness of this technology 
resides in its low radiation hardness. Even if CMOS components are more robust than 
CCD the qualification of these 2D detectors for neutron fluxes of ITER level remains 
an open question and potentially an interesting field of research.

\section{\label{sec:six} Conclusions and ITER prospects}

JET real time control system nowadays includes a wide set of diagnostics, covering 
the magnetic configuration, the current profile and the main kinetic parameters of the 
electron and ion fluid. These measurements are complemented by other relevant tools, 
which provide derived quantities of major interest like the plasma position, shape and 
topology. The implemented multi-platform architecture, based on industry 
transmission standards, combines the desired flexibility with the robustness required 
by JET program.  These new tools were an essential prerequisite of some of the most 
ambitious feedback program at JET. The implementation of the XSC is not only 
extremely relevant for the next years experimental program but was also an unique 
opportunity to test ITER control techniques of the shape, based on calculating the 
plasma response models directly from equilibrium codes. The simultaneous control of 
the current and pressure profile constitutes one of the most significant programs on 
the route to the feedback control of ITBs. New and more advanced approached, based 
on soft computing, were also validated. Innovative technologies are also promoted, 
particularly in the field of real time imaging, which requires more advanced 2D 
sensors for some applications.  From the technological and architectural point of view, 
the present JET real time control system seems therefore to provide a lot of useful 
information not only in support of the experimental program but also for the design of 
ITER. 

The relatively recent but substantial experience gathered on feedback control at JET 
in the last years allows assessing which are the main requirements the diagnostics 
have to fulfil in general to become good candidates for real time. In order to use the 
measurement of a certain physical quantity in real time, an established method to 
measure it must be available. A sound interpretation of the data is essential and the 
information provided needs to be sufficiently closed to the plasma parameter to be 
controlled. In this respect, a good example of a difficult diagnostic to interpret is the 
MSE, if this measurement is to be used to control the $q$ profile. Since the MSE is a 
local measurement whereas the definition of $q$ is an average over a flux surface, 
delicate calculations are needed to derive the correct information for the control from 
the direct measurements \cite{Giannella}. Another delicate aspect of real time diagnostics is 
reliability, which is not limited to the hardware and basic interpretative software but 
has also to take into account possible disturbances from the environment. This means 
that the diagnostic must be robust enough to produce acceptable data even in case of 
major and unforeseen variations of the plasma parameters or the mode of operation 
(ELMs, limiter and divertor configurations etc). The calibration of the diagnostic is 
also a significant issue. An established procedure is necessary, which does not need to 
be carried out necessarily in real time but must be stable enough to guarantee 
meaningful outputs at least for the all discharge. The time constant of the 
measurement technique and the computational time necessary to interpret it have 
obviously also to be compatible with the plasma phenomena to be controlled. Given 
the rate of development of computers, it is very likely that in the perspective of ITER 
silicon technology, particularly if organised in parallel architectures, will be able to 
provide enough computational power even in the case of the most demanding 
diagnostics. On the other hand, for some essential parameters of interest for ITER 
physics and operation, no measurement technique is completely established yet. The 
most delicate field is certainly the one of burning plasma diagnostics. The 
measurements of the isotopic composition, the He ash, the slowing down and lost 
alphas require very significant efforts to identify the most suitable concepts even for 
providing reliable data, letting alone the real time aspects. In the case of neutrons, 
even if quite sound approaches have been tested at JET for the determination of the 
total yield, already available in real time, high-resolution spectrometry is still a 
controversial issue. More work is certainly required in particular to identify solutions, 
which could provide basic information, like the yield of thermal neutrons, with the 
potential of providing $Q_{thermal}$ in real time. Other weaknesses of present day Tokamak 
diagnostics are certainly the measurement of the current density and temperature at 
the edge. For these quantities also a very high time resolution would be necessary, to 
be able to follow edge fast phenomena like the ELMs. One additional category of 
measurements very problematic for ITER are the diagnostics for the divertor. The 
temperature, erosion and redeposition of the divertor plates require significant 
developments of present techniques. Moreover the plasma parameters are going to be 
so extreme in ITER divertor that also the basic measurements of electron density and 
temperature are believed to be very difficult in that environment. In all these fields the 
identification of reliable methods are of course a prerequisite to attack the issue of 
providing the measurements in real time.

It must also be kept in mind that in ITER, a global control system, using all or almost 
all the available actuators (coils, gas injection, coolant, additional heating, tritium 
etc.), could become indispensable. In this respect, a lot of work remains to be done 
not only to significantly develop actuators and sensors, particularly in the direction of 
an increased reliability, but also in devising and testing "integrated" approaches, of 
both adequate complexity and realistic robustness.

\end{document}